\documentclass[aps,prb,showpacs,superscriptaddress]{revtex4}
\usepackage{amsfonts}
\usepackage{txfonts}
\usepackage{epsfig}

\def\be{\begin{equation}} \def\ee{\end{equation}}
\def\bea{\begin{eqnarray}} \def\eea{\end{eqnarray}}

\def\be{\begin{equation}} \def\ee{\end{equation}}
\def\bea{\begin{eqnarray}} \def\eea{\end{eqnarray}}

\begin{document}

\title[Equivalent topological invariants of topological insulators]
{Equivalent topological invariants of topological insulators}
\author{Zhong Wang} 
\affiliation{Department of Modern Physics, University of Science and
Technology of China, Hefei, 230026, P. R. China}
\affiliation{Department of Physics, Stanford University, Stanford,
    CA 94305, USA}
 
\author{Xiao-Liang Qi}

\affiliation{Microsoft Research, Station Q, Elings Hall, University
of California, Santa Barbara, CA 93106, USA}

\affiliation{Department of Physics, Stanford University, Stanford,
    CA 94305, USA}

\author{Shou-Cheng Zhang}

\affiliation{Department of Physics, Stanford University, Stanford,
    CA 94305, USA}

\begin{abstract}
A time-reversal invariant topological insulator can be generally
defined by the effective topological field theory with a quantized
$\theta$ coefficient, which can only take values of $0$ or $\pi$.
This theory is generally valid for an arbitrarily interacting system
and the quantization of the $\theta$ invariant can be directly
measured experimentally. Reduced to the case of a non-interacting
system, the $\theta$ invariant can be expressed as an integral over
the entire three dimensional Brillouin zone. Alternatively,
non-interacting insulators can be classified by topological
invariants defined over discrete time-reversal invariant momenta. In
this paper, we show the complete equivalence between the integral
and the discrete invariants of the topological insulator.
\end{abstract}


\maketitle

\section{Introduction}
In condensed matter systems, most states of matter are classified by
the symmetries they break. For example, a crystal breaks the
translational symmetry, a magnet breaks the rotational symmetry and
a superconductor breaks the gauge symmetry. These broken symmetry
states are identified by the order parameter, and described by the
effective field theory. The effective field theory is constrained by
the broken symmetry, and it defines universality classes of
different states of matter, and predicts universal physical
properties.

The quantum Hall (QH) state is a topological state of matter which
does not fall into the conventional Landau paradigm. The long
distance and low energy properties of the QH state is generally
described by the effective topological field theory
\begin{eqnarray}
S_{\rm eff}&=&\frac{C_1}{4\pi}\int d^2x dt \epsilon^{\mu\nu\tau}
A_\mu\partial_\nu A_\tau,\label{Seff2d} \label{cs2d}
\end{eqnarray}
This topological field theory is generally valid for interacting
systems, and describes the electromagnetic response in the long wave
length limit\cite{zhang1992}. In the non-interacting limit, the Hall
conductance $\sigma_{xy}=C_1$ can be expressed as a topological
invariant over the two-dimensional (2D) Brillouin
zone\cite{thouless1982}, given by
\begin{eqnarray}
C_1&=&\frac{e^2}{h}\frac1{2\pi}\int dk_x \int dk_y f_{xy}\left({\bf
k}\right) \label{1stChern}
\end{eqnarray}
where
\begin{eqnarray} f_{xy}\left({\bf k}\right)&=&\frac{\partial
a_y({\bf k})}{\partial
k_x}-\frac{\partial a_x({\bf k})}{\partial k_y}\nonumber\\
a_i({\bf k})&=&-i\sum_{\alpha\in ~{\rm occ}}\left\langle \alpha{\bf
k}\right|\frac{\partial}{\partial k_i}\left|\alpha{\bf
k}\right\rangle,~i=x,y.\nonumber
\end{eqnarray}
are the abelian Berry curvature and potential for the band states
$|\alpha{\bf k}\rangle$. Time reversal (TR) symmetry breaking is
essential for the QH effect.

More recently, new topological insulator states have been
theoretically predicted and experimentally observed in HgTe quantum
wells, BiSb alloys, Bi2Te3 and Bi2Se3
crystals\cite{bernevig2006d,koenig2007,fu2007a,
hsieh2008,zhang2009,xia2009,chen2009}. These topological insulator
states physically arises from the spin-orbit coupling in electronic
structures, and are protected by the TR
symmetry\cite{murakami2003,qi2008,fu2007b,fu2007a,moore2007}. A
general theory describes the three dimensional (3D) topological
insulator in terms of a effective topological field
theory\cite{qi2008}, given by

\begin{eqnarray}
S_{\rm eff}=S_{\rm Maxwell}+S_{\rm topo}=\int
d^3xdt\left[\frac1{16\pi}F_{\mu\nu}F^{\mu\nu}+\frac{\theta\alpha}{32\pi^{2}}
\epsilon^{\mu\nu\sigma\tau}
F_{\mu\nu}F_{\sigma\tau}\right]\label{Stot3d} \label{theta}
\end{eqnarray}
where the last term is the topological term and $\alpha=e^2/hc$ is
the fine structure constant. For a period system, the partition
function is invariant under the shift of $\theta$ by integer
multiples of $2\pi$. Since TR transformation replaces $\theta$ by
$-\theta$, only two special, quantized values of $\theta=0$ and
$\theta=\pi$ are consistent with the TR symmetry. Therefore, all TR
invariant insulators fall into the two distinct classes,
disconnected from each other. The Standard Model of elementary
particles also admits such a topological term, therefore, the study
of the topological insulator can also shed light on the fundamental
topological interactions in Nature\cite{wilczek2009}.

We can always couple an arbitrarily interacting electron system to
the external electromagnetic field, and integrate out all the
electronic degrees of freedom to obtain the effective field theory
(\ref{theta}). For a TR invariant insulator, only two discrete
possibilities of $\theta=0$ and $\theta=\pi$ can be realized. For a
non-interacting band insulator, Qi, Hughes and Zhang (QHZ) gave an
explicit formula for the $\theta$ parameter as an integral over the
3D Brillouin zone
\begin{eqnarray}
\theta \equiv 2\pi P_{3}(\theta) = \frac{1}{8\pi} \int d^{3}{\bf k}
\epsilon^{ijk} \textrm{Tr}\{[f_{ij}({\bf k})-\frac{2}{3}i a_{i}({\bf
k})\cdot a_{j}({\bf k})]\cdot a_{k}({\bf k})\} \label{qhz}
\end{eqnarray}
where
\begin{eqnarray} f^{\alpha\beta}_{ij}&=&\partial_i
a^{\alpha\beta}_j-\partial_j
a^{\alpha\beta}_i+i\left[a_i,a_j\right]^{\alpha\beta},\nonumber\\
a_i^{\alpha\beta}({\bf k})&=&-i\left\langle \alpha,{\bf
k}\right|\frac{\partial}{\partial k_i }\left|\beta,{\bf
k}\right\rangle\nonumber
\end{eqnarray}
are the non-abelian Berry curvature and potential for the band state
$\left|\beta,{\bf k}\right\rangle$. This topological invariant has
the physical interpretation of a magneto-electric polarization,
which can be directly measured experimentally\cite{qi2008,qi2009}.

We notice a beautiful symmetry between the topological field theory
of the TR breaking QH state and the TR invariant topological
insulator state. Both the space-time integrals (\ref{cs2d}),
(\ref{theta}) and the Brillouin zone integrals (\ref{1stChern}),
(\ref{qhz}) are integral topological invariants in the theory of
differential geometry. We see that the TR breaking QH state is
described by a Chern-Simons integral over the $2+1$ dimensional
space-time (\ref{cs2d}), and a first Chern integral over the 2D
Brillouin zone (\ref{1stChern}), while the TR invariant topological
insulator is described by the second Chern integral over the $3+1$
dimensional space-time (\ref{theta}) and a Chern-Simons integral
(\ref{qhz}) over the 3D Brillouin zone. These are the deepest and
most natural topological invariants in mathematics, and it is
gratifying to see that they also describe topological states
realized in Nature.

TR invariant insulators form an universality class extending over
the 4D, 3D and 2D space. The root state of this universality class
is the topological insulator in 4D\cite{zhang2001,bernevig2002}. In
fact, it was the first TR invariant insulator state introduced
theoretically, and historically it was referred to as the 4D QH
state. It is described by a topological field theory
\begin{eqnarray}
S_{\rm eff}=\frac{C_2}{24\pi^2}\int
d^4xdt\epsilon^{\mu\nu\rho\sigma\tau}A_\mu\partial_\nu
A_\rho\partial_\sigma A_\tau \label{Seff4d}
\end{eqnarray}
Under TR transformation,
\begin{eqnarray}
A_0 \rightarrow A_0  \label{TR}, \, \, A_{i} \rightarrow -A_{i}
\end{eqnarray}
therefore, we see that the topological field theory in $2+1$
dimensions (\ref{cs2d}) breaks TR symmetry, whereas the topological
field theory in $4+1$ dimensions (\ref{Seff4d})  preserves the TR
symmetry, and naturally describes the TR invariant topological
insulators. For the case of non-interacting fermions, the
coefficient is given explicitly by the second Chern number
\begin{eqnarray}
C_2&=&\frac1{32\pi^2}\int d^4k\epsilon^{ijk\ell}{\rm
tr}\left[f_{ij}f_{k\ell}\right] \label{2chern}
\end{eqnarray}

The TR invariant topological insulator defined by (\ref{Seff4d}) and
(\ref{2chern}) in the 4D space naturally generalizes the
corresponding quantities of the TR breaking topological insulator
defined by (\ref{cs2d}) and (\ref{1stChern}) in the 2D space. This
is the reason why it was historically referred to as the 4D QH
state\cite{zhang2001,bernevig2002}. It is the root state for all TR
invariant topological insulators in 3D and 2D, which can be obtained
from the root state in 4D through the process of dimensional
reduction\cite{qi2008}.

Topological invariants in 3D band insulators have been studied from
a different approach\cite{fu2007b,moore2007}. In a beautiful series
of papers, Fu, Kane and Mele (FKM) introduced a $Z_2$ topological
invariant\cite{fu2006, fu2007a, fu2007b} for the strong topological
insulator, expressed as a discrete product over the eight
time-reversal invariant momenta (TRIM), explicitly written as
\begin{eqnarray}
(-1)^{\nu_{0}}=\prod_{i=1}^{8}\delta_{i}  \label{Z2}
\end{eqnarray}
where $\delta_{i}$ is quantity defined at TRIM $\Gamma_{i}$, the
detailed form of which will be given in the following sections.
Since weak topological insulators may not be generally robust, we
shall not be concerned with their definitions here.

Therefore, there are now two topological invariants defined for the
topological insulator. They are motivated by different logical
reasoning and have different mathematical forms. The integral
invariant (\ref{qhz}) given by QHZ is physically measurable in terms
of the magneto-electric polarization, and leads directly to the
general topological field theory (\ref{theta}). The discrete
invariant (\ref{Z2}) given by FKM has the distinct advantage that it
can be easily evaluated, especially for crystals with inversion
symmetry. Applied to concrete models of topological
insulators\cite{qi2008,essin2009}, these two definitions yield the
same result. However, it is highly desirable to prove the general
equivalence between these two definitions. In this paper, we show
explicitly that one can transform the integral invariant (\ref{qhz})
to the discrete invariant (\ref{Z2}) exactly, proving the precise
equivalence between these two definitions.

\section{A pedagogical example}
Because our approach involved the mathematical concept of degree of
map, we shall give a brief introduction to it. Our presentation in
this section is mainly designed for intuitive understanding, rather
than mathematical rigor. Therefore, we shall illustrate the idea of
degree of map in a simple one-dimensional example. For more
mathematical details, {\it c. f.} Ref. \cite{dubrovin1985}. Those
readers who are already familiar with this subject may skip this
section.

\begin{figure}
\includegraphics[width=15cm]{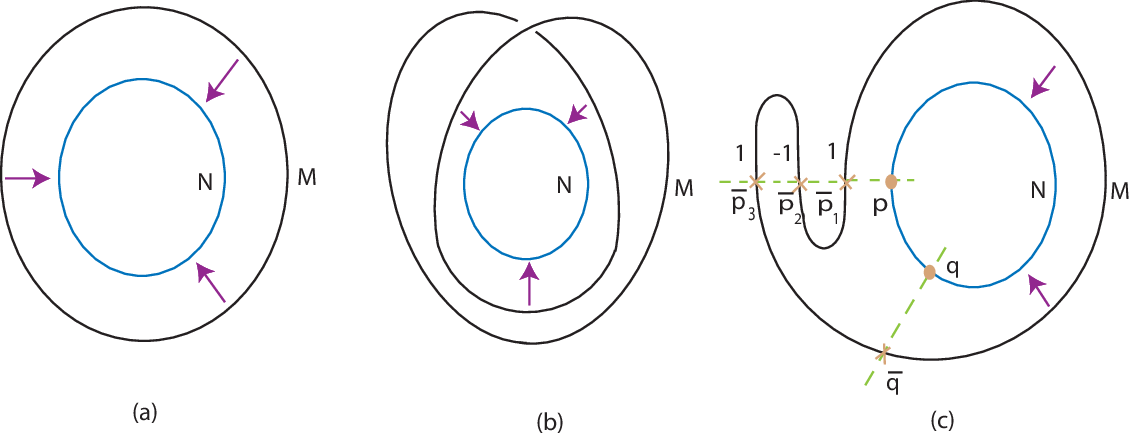} \caption{Map
between two circles $M$ and $N$. The arrows indicate the map. The
winding number of (a), (b) and (c) is 1,2 and 1 respectively. }
\label{degree}
\end{figure}

Let us consider one concrete example. Consider a map $f: \, M
\rightarrow N$, where $M$ and $N$ are both one-dimensional circle
$S^{1}$ (see Fig.\ref{degree}). The coordinates of $M$ and $N$ are
denoted as $\phi$ and $\theta$, respectively. The degree of map $f$
is just the number of times that $M$ covers $N$ under $f$ and it is
often called ``winding number".

The standard integral form of the winding number is defined as
\begin{eqnarray}
\textrm{deg}(f) =
\frac{1}{2\pi}\int_{\phi=0}^{\phi=2\pi}d\theta(\phi) =
\frac{1}{2\pi}\int_{\phi=0}^{\phi=2\pi}\frac{d\theta}{d\phi} d\phi =
n \in \textrm{Z}
\end{eqnarray}
which has a simple geometrical interpretation. For simplicity, let
us assume that the point $\phi=0$ maps to $\theta=0$. When $\phi$
goes from $0$ to $2\pi$, $\theta$ goes from $0$ to $2\pi n$. This
integral form of the winding number can be expressed in a discrete
form. We arbitrarily choose a image point $p$ on the image manifold
$N$ (see Fig.\ref{degree} for illustration), and count the number of
source points $\overline{p_1}$, $\overline{p_2}$, $\overline{p_3}$,
etc on the source manifold $M$, where the source points are weighted
by the $+1$ or $-1$ sign depending on the orientation of the map.
For example, image point $q$ has only one source point
$\overline{q}$, which maps onto $q$ in a clockwise sense. Therefore
$\textrm{deg}(f) =1$. On the other hand, the image point $p$ has
three source points $\overline{p_1}$, $\overline{p_2}$ and
$\overline{p_3}$. $\overline{p_1}$ and $\overline{p_3}$ map onto $p$
in a clockwise sense, while $\overline{p_2}$ maps onto $p$ in a
counterclockwise sense, giving $\textrm{deg}(f) =+1-1+1=1$. We see
from this example that the integral invariant of the winding number
can be reduced to counting the weighted number of source points of a
given image point -- this is a discrete invariant for the winding
number.

The above picture can be generalized to higher dimensions.
Generally, for a map $f: \, M \rightarrow N$, where $M$ and $N$ are
both $d$ dimensional orientable manifold, we can defined the degree
of map as
\begin{eqnarray}
\textrm{deg}(f) = \int_{M}f^{*}(\omega)\label{integral}
\end{eqnarray}
where $\omega$ is a $d-$form volume element on $N$ satisfying
$\int_{N}\omega=1$, and $f^{*}(\omega)$ is the pullback of $\omega$
to $M$ under $f$. In our one-dimensional example discussed earlier,
$\omega=d\theta/2\pi$, and
$f^{*}(\omega)=d\theta(\phi)/2\pi=(d\theta/d\phi)d\phi/2\pi$.

In our one dimensional example, we have seen that there is a
discrete form of the invariant $\textrm{deg}(f)$. This can also be
generalized to higher dimension. It is given as\cite{dubrovin1985}
\begin{eqnarray}
\textrm{deg}(f) = N[f^{-1}(p), \, J_{f^{-1}(p)}>0]- N[f^{-1}(p), \,
J_{f^{-1}(p)}<0] \label{discrete}
\end{eqnarray}
where $J$ is the Jacobian of the map, and $N[f^{-1}(p),\,
J_{f^{-1}(p)}>(<)0]$ denotes the number of source points mapping to
$p$ with a positive (negative) Jacobian, respectively. Without going
into the general derivation of this formula, we just give the
explanation in our one dimensional example. In that example,
$J_{f^{-1}(p)}$ is the direction sense (clockwise or
counterclockwise) of the map at the point $p$, which is exactly what
we have discussed earlier. We also mention that the point $p$ in
Eq.(\ref{discrete}) should be regular, which means that the Jacobian
of map are nonzero at $f^{-1}(p)$. Because the set of non-regular
points has zero measure, we can always perturb the map to remove
non-regularity at a given point.

In summary, we showed that the degree of a map can be expressed in
two equivalent forms, the integral form (\ref{integral}) and the
discrete form (\ref{discrete}). In the next section, we will apply
this idea to the integral and the discrete invariants of (\ref{qhz})
and (\ref{Z2}), which are degree of map $\textrm{deg}(f)$ modulo
$2$, simply denoted as $\textrm{deg}_{2}(f)$.

\section{Equivalence between integral and discrete topological invariants}
Let us start from the band structure of time-reversal invariant
topological insulators. To simplify the problem, we first assume
that there are no degeneracies except those required by time
reversal symmetry. Suppose that there are $2N$ filled bands labeled
by $(\alpha,\beta,\cdots)$, where the number of filled bands is even
because of the Kramers degeneracy. The $2N$ by $2N$ matrix $B({\bf
k})$ is defined by
\begin{eqnarray}
| -{\bf k},\alpha \rangle = \sum_{\beta} B^{\ast}_{\alpha
\beta}({\bf k}) | \Theta, {\bf k}, \beta \rangle
\end{eqnarray}
where $ | \Theta, {\bf k}, \beta \rangle = \hat{\Theta} |{\bf
k},\beta \rangle$ and $\hat{\Theta}$ is the TR operator. The
property $\hat{\Theta}^{2}=-1$ is crucial for TR invariant Fermi
systems. By direct calculation we get the inner product
\begin{eqnarray}
\langle -{\bf k},\alpha | \Theta, {\bf k},\beta \rangle &=&
\sum_{\gamma}B_{\alpha \gamma}({\bf k}) \langle \Theta, {\bf k},
\gamma | \Theta,
{\bf k}, \beta \rangle \nonumber \\
&=& \sum_{\gamma}B_{\alpha \gamma}({\bf k}) \delta_{\gamma \beta}
\nonumber \\
&=& B_{\alpha \beta}({\bf k})
\end{eqnarray}
and an important property of $B({\bf k})$
\begin{eqnarray}
B_{\alpha \beta} (-{\bf k}) &=& \langle {\bf k},\alpha | \Theta,
-{\bf k}, \beta
\rangle \nonumber \\
&=& - \langle -{\bf k},\beta | \Theta, {\bf k}, \alpha \rangle
\nonumber\\
&=& -B_{\beta \alpha}({\bf k})
\end{eqnarray}
where the facts that $\hat{\Theta}^{2} = -1$ and $\langle \alpha |
\Theta, \beta \rangle = - \langle \beta | \Theta, \alpha \rangle $
have been used. Therefore, $B_{\alpha \beta} (-{\bf k})$ is
anti-symmetric at the eight TR invariant momenta(TRIM), which
enables the definition of Pfaffian at these points. The definition
of the discrete invariant given by FKM\cite{fu2006,fu2007b} is
expressed as
\begin{displaymath}
(-1)^{\nu_{0}} = \prod_{i=1}^{8} \delta _{i}\ \ \ ;\ \ \ \delta_{i}
= \frac{ \sqrt{\textrm{det}[B(\Gamma_{i})]} }{
\textrm{Pf}[B(\Gamma_{i})] } = \pm 1
\end{displaymath}
with the Pfaffian of $B$ written as $\textrm{Pf}[B]$. Although it
seems that all the quantities appearing in this definition are
local, global information on the Brillouin zone
$T^{3}$(three-dimensional torus) is encoded because a global basis
of wavefunction is required in this definition. The existence of
global basis is not obvious and we shall present a discussion on
this in the appendix. The unitary matrix $B({\bf k})$ defines a map
\begin{eqnarray}
f: T^{3} \rightarrow U(2N) \end{eqnarray} This map can be simplified
since we have the assumption of absence of accidental degeneracies.
We divide the $2N$ filled bands into $N$ TR pairs. Because energy
eigenstates with different eigenenergy are orthogonal, we have
$B_{\alpha \beta}({\bf k})=0$ if $|-{\bf k},\alpha \rangle$ and
$|{\bf k},\beta \rangle $ belong to different pairs. Therefore, all
the inter-pairs elements of matrix $B({\bf k})$ are zero. Written
explicitly, the matrix $B({\bf k})$ takes the following form
\begin{eqnarray}
B({\bf k}) = \left[
                \begin{array}{ccccc}
                  B_{1}({\bf k}) &  &  &  &  \\
                   & B_{2}({\bf k}) &  & &  \\
                   & & B_{3}({\bf k}) &  &  \\
                   &  &  & \ddots &  \\
                   &  & & & B_{N}({\bf k}) \\
                \end{array}
 \right]
\end{eqnarray}
where each $B_{m}({\bf k})$ is an $U(2)$ matrix. Therefore, the map
$f$ splits into $N$ maps
\begin{eqnarray}
f_{m} : T^{3} \rightarrow U(2) \qquad m = 1,2,\cdots,N
\end{eqnarray}

\begin{figure}
\includegraphics[width=13cm]{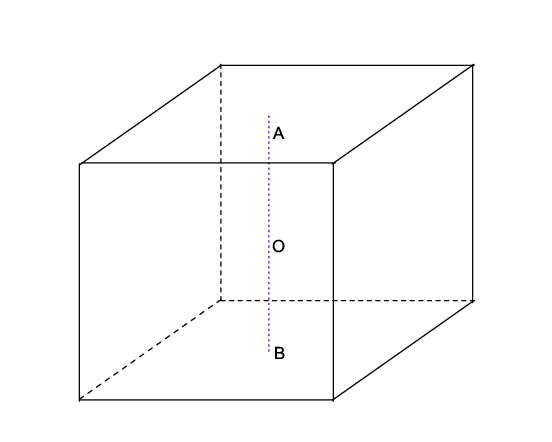} \caption{The
loop (closed line) $AOB$ in $T^{3}$. The point $O$ is $k=(0,0,0)$.
Points $A$ and $B$ are the same point because the periodic boundary
condition. $AOB$ is homotopically nontrivial, \emph{i.e.} $AOB$
cannot be continuously deformed to a single point.} \label{t3}
\end{figure}

The next step is reducing $U(2)$ to $SU(2)$, which is a
three-dimensional manifold. The motivation of this step will be
clear later when we consider the degree of certain map. To this end,
we factorize $B_{m}({\bf k})= \exp[i\theta_{m}({\bf k})] u_{m}({\bf
k})$, where $u_{m}({\bf k})$ is a $SU(2)$ matrix function of ${\bf
k}$. Because of the facts that $\exp[i\theta_{m}({\bf k})]
u_{m}({\bf k})=-\exp[i\theta_{m}({\bf k})] [-u_{m}({\bf k})]$ and
$-u_{m}({\bf k})$ is also $SU(2)$ matrix, there is ambiguity in this
factorization. To avoid this ambiguity, we can choose a point ${\bf
k}_{0}$ in $T^{3}$, and choose one factorization $B_{m}({\bf
k}_{0})= \exp[i\theta_{m}({\bf k}_{0})] u_{m}({\bf k}_{0})$. The
$U(1)$ factors of other ${\bf k}$ points are determined by

\begin{eqnarray}
\theta_{m}({\bf k})= \theta_{m}({\bf k}_{0}) -\frac{i}{2} \int_{{\bf
k}_{0}}^{{\bf k}}d {\bf k} \nabla_{{\bf
k}}\ln(\textrm{det}[B_{m}({\bf k})])
\end{eqnarray}
It is easy to check that with this equation $u_{m}(k)$ will have
unitary determinant. To make the integral in the above equation
unambiguous, we require that $\oint_{l}d{\bf k} \nabla_{{\bf
k}}\textrm{det}[B_{m}({\bf k})] = 0$ for arbitrary loop $l$ in
$T^{3}$. For contractable loops, this is trivially satisfied. For
non-contractable loop, consider the non-contractable loop $AOB$ in
Fig.\ref{t3} as an example, we have

\begin{eqnarray}
\oint_{AOB}d {\bf k} \nabla_{{\bf k}}\ln(\textrm{det}[B_{m}({\bf
k})]) &=& \int_{AO}d {\bf k} \nabla_{{\bf
k}}\ln(\textrm{det}[B_{m}({\bf k})]) + \int_{OB}d {\bf k}
\nabla_{{\bf k}}\ln(\textrm{det}[B_{m}({\bf k})]) \nonumber\\ &=& 0
\end{eqnarray}
which is a result of the relation $\textrm{det}[B_{m}(-{\bf
k})]=\textrm{det}[B_{m}({\bf k})]$. Therefore, we have a
factorization of $U(2)$ into $SU(2)$ and $U(1)$. With this
factorization, we also have that $\oint_{l}d {\bf k} \nabla_{{\bf
k}} \theta_{m}({\bf k}) = 0$ (for arbitrary loop $l$). Therefore, we
can adiabatically deform the $U(1)$ factor to $1$. Having got rid of
the $U(1)$ factor, the maps $f_{m}: T^{3} \rightarrow U(2)$ are
deformed to $g_{m}: T^{3} \rightarrow SU(2)$. This completes our
discussion on matrix $B({\bf k})$. With the aid of $B({\bf k})$, we
can give geometrical interpretation of topological invariants of TR
invariant topological insulators.

In order to prove the equivalence, we first show how to relate the
topological invariant $P_3$ in Eq. (\ref{qhz}) to a winding number.
Similar formula appeared in Ref.\cite{qi2008} in a slightly
different form but we would like to include the discussion here to
make our discussion self-contained. We start from the integral
topological invariants
\begin{eqnarray} P_{3} = \frac{1}{16\pi^{2}} \int
d^{3}{\bf k}\epsilon^{ijk} \textrm{Tr}\{[f_{ij}({\bf
k})-\frac{2}{3}i a_{i}({\bf k})\cdot a_{j}({\bf k})]\cdot a_{k}({\bf
k})\}
\end{eqnarray}
which is valid independent of the TR symmetry. In a TR invariant
system, with the aid of $B({\bf k})$, we have
\begin{eqnarray}
a_{i}^{\alpha \beta}(-{\bf k}) &=& -i \langle -{\bf k},\alpha |
\partial_{-{\bf k}_{i}} |-{\bf k},\beta \rangle   \nonumber \\
&=& i \sum_{\alpha'\beta'} B_{\alpha \alpha'} \langle \Theta,{\bf
k},\alpha' | \partial_{{\bf k}_{i}} (B_{\beta \beta'}^{\ast}
|\Theta,{\bf k},\beta' \rangle )  \nonumber \\
&=& i \sum_{\alpha'\beta'} B_{\alpha \alpha'} B_{\beta
\beta'}^{\ast} \langle \Theta,{\bf k},\alpha' | \partial_{{\bf
k}_{i}} |\Theta,{\bf k},\beta' \rangle + i \sum_{\alpha'\beta'}
B_{\alpha \alpha'}
\partial_{{\bf k}_{i}} B_{\beta \beta'}^{\ast} \delta_{\alpha'\beta'}
\nonumber \\
&=& \sum_{\alpha'\beta'} B_{\alpha \alpha'}
(a_{i}^{\alpha'\beta'}({\bf k}))^{\ast} B_{\beta'\beta}^{\dagger} +
i\sum_{\alpha'} B_{\alpha \alpha'}\partial_{{\bf k}_{i}} B_{\alpha'
\beta}^{\dagger}
\end{eqnarray}
which can be simply written as
\begin{eqnarray} a_{i}(-{\bf k}) = B({\bf k}) a_{i}^{\ast}({\bf k})
B^{\dagger}({\bf k}) + i B({\bf k}) \partial_{i}B^{\dagger}({\bf k})
\end{eqnarray}
Therefore, the field strength satisfies the following relation
\begin{eqnarray}
f_{ij}(-{\bf k}) &=& \partial_{-{\bf k}_{i}} a_{j}(-{\bf k}) -
\partial_{-{\bf k}_{j}}
a_{i}(-{\bf k}) + i[a_{i}(-{\bf k}),a_{j}(-{\bf k})] \nonumber \\
&=& -B({\bf k})f_{ij}^{\ast}({\bf k})B^{\dagger}({\bf k})
\end{eqnarray}
These results imply that the non-abelian Berry potential and
curvature at $\bf -k$ are simply related to those at $\bf k$ by a
non-abelian gauge transformation $B({\bf k})$. Therefore, we obtain
that
\begin{eqnarray}
P_{3} &=& \frac{1}{16\pi^{2}} \int d^{3}{\bf k}\epsilon^{ijk}
\textrm{Tr}\{[f_{ij}(-{\bf k})-\frac{2}{3}i a_{i}(-{\bf
k})a_{j}(-{\bf k})]a_{k}(-{\bf k})\}
\nonumber \\
&=& -\frac{1}{16\pi^{2}} \int d^{3}{\bf k}\epsilon^{ijk}
\textrm{Tr}\{[f_{ij}({\bf k})-\frac{2}{3}i a_{i}({\bf k})a_{j}({\bf
k})]a_{k}({\bf k})\}^{\ast} \nonumber \\ && - \frac{i}{8\pi^{2}}
\int d^{3}{\bf k}\epsilon^{ijk} \textrm{Tr}\partial_{i}(B
a_{j}^{\ast}\partial_{k} B^{\dagger}) \nonumber \\
&&- \frac{1}{24\pi^{2}} \int d^{3}{\bf k} \epsilon ^{ijk}
\textrm{Tr}[(B\partial_{i}B^{\dagger})( B\partial_{j}B^{\dagger})(
B\partial_{k}B^{\dagger} )]
\nonumber \\
&=& -P_{3}^{\ast} - \frac{1}{24\pi^{2}} \int d^{3}{\bf k} \epsilon
^{ijk} \textrm{Tr}[(B\partial_{i}B^{\dagger})(
B\partial_{j}B^{\dagger})( B\partial_{k}B^{\dagger} )]
\nonumber \\
&=& -P_{3}- \frac{1}{24\pi^{2}} \int d^{3}{\bf k} \epsilon ^{ijk}
\textrm{Tr}[(B\partial_{i}B^{\dagger})( B\partial_{j}B^{\dagger})(
B\partial_{k}B^{\dagger} )]
\end{eqnarray}
or
\begin{eqnarray}
2P_{3} = -\frac{1}{24\pi^{2}} \int d^{3}{\bf k} \epsilon ^{ijk}
\textrm{Tr}[(B\partial_{i}B^{\dagger})( B\partial_{j}B^{\dagger})(
B\partial_{k}B^{\dagger} )]
\end{eqnarray}
Both the LHS and RHS of the above equation depend on gauge choices,
but the parity of $2P_3$ is gauge invariant
\begin{eqnarray}
2P_{3}(\textrm{mod}\, 2) = -\frac{1}{24\pi^{2}} \int d^{3}{\bf k}
\epsilon ^{ijk} \textrm{Tr}[(B\partial_{i}B^{\dagger})(
B\partial_{j}B^{\dagger})( B\partial_{k}B^{\dagger} )] \,
(\textrm{mod}\, 2)
\end{eqnarray}
which is the important quantity for characterization of the
topological insulator.

In our previous discussion, we have split $B(k)$ into direct sum of
$SU(2)$ matrices($B_{m},m=1,2,\cdots,N$). It is readily obtained
that
\begin{eqnarray}
2P_{3}(\textrm{mod}\, 2) = \sum_{m=1}^{N}\nu_{m} \quad
(\textrm{mod}\, 2)
\end{eqnarray}
where
\begin{eqnarray}
\nu_{m}=  -\frac{1}{24\pi^{2}} \int d^{3}{\bf k} \epsilon ^{ijk}
\textrm{Tr}[(B_{m}\partial_{i}B_{m}^{\dagger})(
B_{m}\partial_{j}B_{m}^{\dagger})( B_{m}\partial_{k}B_{m}^{\dagger}
)]
\end{eqnarray}
We notice that $\nu_{m}$ is just the winding number, or the integral
form of degree of the map
\begin{eqnarray}
g_{m}:\,T^{3} \rightarrow SU(2)
\end{eqnarray}
The mod 2 degree of this map is given by
\begin{eqnarray}
 \textrm{deg}_{2}(g_{m})= \nu_{m}(\textrm{mod} \, 2)
\end{eqnarray}
Therefore we have
\begin{eqnarray}
 2P_{3}(\textrm{mod}\, 2) = \sum_{m=1}^{N} \textrm{deg}_{2}(g_{m})
\end{eqnarray}

This is the integral form of degree of map. The natural question is
whether we can find a discrete form, as discussed in the previous
section. Let us denote the image of ${\bf k}$ under the map $g_{m}$
as $B_{m}({\bf k})$, which is a $SU(2)$ matrix. As has been noted,
$B_{m}(\Gamma_{i})$ is anti-symmetric at TRIM $\Gamma_{i}$. There
are only two anti-symmetric matrices in $SU(2)$, given by
\begin{displaymath}
A_{1} = \left(
          \begin{array}{cc}
            0 & 1 \\
            -1 & 0 \\
          \end{array}
        \right),
        \qquad
A_{2} = \left(
          \begin{array}{cc}
            0 & -1 \\
            1 & 0 \\
          \end{array}
        \right)
\end{displaymath}
It is easy to see that $\textrm{Pf}[A_{1}]=1$ and
$\textrm{Pf}[A_{2}]=-1$. In principle, we can pick any image point
on the $SU(2)$ to perform the counting for the degree of the map.
However, because of the TR symmetry, the counting is particularly
simple if we pick either $A_{2}$ or $A_{1}$. We now count the number
of source points ${\bf k}$, modulo $2$, which map onto $A_{2}$. We
first notice that if $B_{m}({\bf k})=A_{2}$, then $B_{m}(-{\bf
k})=-B_{m}^{T}({\bf k}) = A_{2}$. Therefore, if ${\bf k}$ maps to
$A_{2}$, so does $-{\bf k}$. If ${\bf k}$ is not one of the TRIM,
${\bf k}$ and ${\bf -k}$ are distinct, and these two points do not
contribute to $\textrm{deg}_{2}(g_{m})$, which counts the number of
source points modulo $2$. The only source points which may
contribute to $\textrm{deg}_{2}(g_{m})$ are TRIM, where ${\bf k}$
and $-{\bf k}$ are equivalent. It is also important to note that the
calculation of mod 2 degree of map is easier than that of the
integer degree, which involves the $\pm$ signs of the Jacobian of
the map. Because $-1=1(\textrm{mod} \, 2)$, we do not need to be
concerned about the signs of the Jacobian when calculating degree of
map mod 2.

Because we have reduced the map from $U(2)$ to $SU(2)$,
$\textrm{det}[B(\Gamma_{i})]=1$, we have $\delta_{i} =
\sqrt{\textrm{det}[B(\Gamma_{i})]}/\textrm{Pf}[B(\Gamma_{i})] =
\textrm{Pf}[B(\Gamma_{i})] = \prod_{m}
\textrm{Pf}[B_{m}(\Gamma_{i})]$. Suppose that there are $n_{m}$
TRIMs which map onto $A_{2}$ under the map $g_{m}$ and $8-n_{m}$
TRIMs which map onto $A_{1}$. By counting the number of points which
map to $A_{2}$ mod $2$, we have $ \textrm{deg}_{2}(g_{m}) =
 n_{m}(\textrm{mod}\, 2)$, and therefore
 \begin{eqnarray}
 (-1)^{2P_{3}} =(-1)^{ \sum_{m=1}^{N}
 \textrm{deg}_{2}(g_{m})} = \prod_{m}(-1)^{n_{m}}
\end{eqnarray}
On the other hand, the discrete invariant defined in Ref.
\cite{fu2007b} is given by
\begin{eqnarray}
(-1)^{\nu_{0}} &=& \prod_{i} \textrm{Pf}[B(\Gamma_{i})] =
\prod_{i,m} \textrm{Pf}[B_{m}(\Gamma_{i})] = \prod_{m}(-1)^{n_{m}}
\end{eqnarray}
Therefore, we proved the central result of this paper, namely the
exact equivalence between the integral invariant of QHZ and the
discrete invariant of FKM for the 3D topological insulator:
\begin{eqnarray}
(-1)^{2P_{3}} = (-1)^{\nu_{0}}
\end{eqnarray}

In the proof given above, we made the assumption that no accidental
degeneracy except the Kramers degeneracy occurs, which simplified
the discussion. The extension to the generic case is straightforward
because $\pi_{3}(U(2N))= \pi_{3}(U(2))$ holds for any integer $N\geq
1$. One can always deform the map $f$ to one of the $U(2)$ subgroups
of $U(2N)$, so that the proof discussed above applies.

We thank Shao-Long Wan and Yong-Shi Wu for helpful
discussions. This work is supported by the US Department of Energy,
Office of Basic Energy Sciences under contract DE-AC03-76SF00515. Z.
Wang acknowledges the support of China Scholarship Council and NSF
of China(Grant No.10675108).

\appendix

\section{Discussion on the global basis of wavefunctions}
In this appendix we shall show that the global basis of wavefunction
exists on $T^{3}$ at presence of time-reversal symmetry. Because the
global definition of $B(k)$ and the discrete invariant $\nu_{0}$
depends on the existence of such global basis, this point is
important to our argument.

We divide the $2N$ filled bands into $N$ TR pairs. Within each pair,
the Hilbert space is two dimensional at each momentum $k$.
Therefore, the Hilbert space is naturally an $U(2)$ fibre bundle on
$T^{3}$. Generically it is not evident that this bundle is trivial.
The key to the existence of global basis is the TR symmetry.

The ``surface'' of the cubic in Fig.\ref{t3} consists of three
$T^{2}$ (two-dimensional tori).  We consider the restriction of the
$U(2)$ bundle to one of these tori. The bundle on $T^{2}$ is trivial
because the first chern number $C_{1}=0$, which is a consequence of
TR symmetry. Therefore, a global basis exists on the surface of the
cubic. Such a basis can be extrapolated to the interior of the cubic
since the the interior region is topologically trivial. We consider
the transition function between the surface and the interior region
on their overlapping region, which is topologically equivalent to
$S^{2}$(two-dimensional sphere). Because the second homotopy group
$\pi_{2}(U(2))$ is trivial, mapping from $S^{2}$ to $U(2)$ (the
structure group of the fibre bundle) are all trivial. Thus the
bundle is trivial on $T^{3}$. Therefore, global basis does exist.

For comparison, we note that in the case of integer quantum Hall
states, the global basis does not exist on the Brillouin zone
$T^{2}$, because the relevant fibre bundle on $T^{2}$ is
non-trivial. From this example we also see that TR symmetry is
necessary for our argument.

\bibliography{P3Z2}

\begin{thebibliography}{20}
\expandafter\ifx\csname natexlab\endcsname\relax\def\natexlab#1{#1}\fi
\expandafter\ifx\csname bibnamefont\endcsname\relax
  \def\bibnamefont#1{#1}\fi
\expandafter\ifx\csname bibfnamefont\endcsname\relax
  \def\bibfnamefont#1{#1}\fi
\expandafter\ifx\csname citenamefont\endcsname\relax
  \def\citenamefont#1{#1}\fi
\expandafter\ifx\csname url\endcsname\relax
  \def\url#1{\texttt{#1}}\fi
\expandafter\ifx\csname urlprefix\endcsname\relax\def\urlprefix{URL }\fi
\providecommand{\bibinfo}[2]{#2}
\providecommand{\eprint}[2][]{\url{#2}}

\bibitem[{\citenamefont{Zhang}(1992)}]{zhang1992}
\bibinfo{author}{\bibfnamefont{S.~C.} \bibnamefont{Zhang}},
  \bibinfo{journal}{Int. J. Mod. Phys. B} \textbf{\bibinfo{volume}{6}},
  \bibinfo{pages}{25} (\bibinfo{year}{1992}).

\bibitem[{\citenamefont{Thouless et~al.}(1982)\citenamefont{Thouless, Kohmoto,
  Nightingale, and den Nijs}}]{thouless1982}
\bibinfo{author}{\bibfnamefont{D.~J.} \bibnamefont{Thouless}},
  \bibinfo{author}{\bibfnamefont{M.}~\bibnamefont{Kohmoto}},
  \bibinfo{author}{\bibfnamefont{M.~P.} \bibnamefont{Nightingale}},
  \bibnamefont{and} \bibinfo{author}{\bibfnamefont{M.}~\bibnamefont{den Nijs}},
  \bibinfo{journal}{Phys. Rev. Lett.} \textbf{\bibinfo{volume}{49}},
  \bibinfo{pages}{405} (\bibinfo{year}{1982}).

\bibitem[{\citenamefont{\textrm{B. A. Bernevig}
  et~al.}(2006)\citenamefont{\textrm{B. A. Bernevig}, \textrm{T. L. Hughes},
  and \textrm{S.C. Zhang}}}]{bernevig2006d}
\bibinfo{author}{\bibnamefont{\textrm{B. A. Bernevig}}},
  \bibinfo{author}{\bibnamefont{\textrm{T. L. Hughes}}}, \bibnamefont{and}
  \bibinfo{author}{\bibnamefont{\textrm{S.C. Zhang}}},
  \bibinfo{journal}{Science} \textbf{\bibinfo{volume}{314}},
  \bibinfo{pages}{1757} (\bibinfo{year}{2006}).

\bibitem[{\citenamefont{K\"onig et~al.}(2007)\citenamefont{K\"onig, Wiedmann,
  Br\"une, Roth, Buhmann, Molenkamp, Qi, and Zhang}}]{koenig2007}
\bibinfo{author}{\bibfnamefont{M.}~\bibnamefont{K\"onig}},
  \bibinfo{author}{\bibfnamefont{S.}~\bibnamefont{Wiedmann}},
  \bibinfo{author}{\bibfnamefont{C.}~\bibnamefont{Br\"une}},
  \bibinfo{author}{\bibfnamefont{A.}~\bibnamefont{Roth}},
  \bibinfo{author}{\bibfnamefont{H.}~\bibnamefont{Buhmann}},
  \bibinfo{author}{\bibfnamefont{L.}~\bibnamefont{Molenkamp}},
  \bibinfo{author}{\bibfnamefont{X.-L.} \bibnamefont{Qi}}, \bibnamefont{and}
  \bibinfo{author}{\bibfnamefont{S.-C.} \bibnamefont{Zhang}},
  \bibinfo{journal}{Science} \textbf{\bibinfo{volume}{318}},
  \bibinfo{pages}{766} (\bibinfo{year}{2007}).

\bibitem[{\citenamefont{Fu and Kane}(2007)}]{fu2007a}
\bibinfo{author}{\bibfnamefont{L.}~\bibnamefont{Fu}} \bibnamefont{and}
  \bibinfo{author}{\bibfnamefont{C.~L.} \bibnamefont{Kane}},
  \bibinfo{journal}{Phys. Rev. B} \textbf{\bibinfo{volume}{76}},
  \bibinfo{eid}{045302} (\bibinfo{year}{2007}).

\bibitem[{\citenamefont{Hsieh et~al.}(2008)\citenamefont{Hsieh, Qian, Wray,
  Xia, Hor, Cava, and Hasan}}]{hsieh2008}
\bibinfo{author}{\bibfnamefont{D.}~\bibnamefont{Hsieh}},
  \bibinfo{author}{\bibfnamefont{D.}~\bibnamefont{Qian}},
  \bibinfo{author}{\bibfnamefont{L.}~\bibnamefont{Wray}},
  \bibinfo{author}{\bibfnamefont{Y.}~\bibnamefont{Xia}},
  \bibinfo{author}{\bibfnamefont{Y.~S.} \bibnamefont{Hor}},
  \bibinfo{author}{\bibfnamefont{R.~J.} \bibnamefont{Cava}}, \bibnamefont{and}
  \bibinfo{author}{\bibfnamefont{M.~Z.} \bibnamefont{Hasan}},
  \bibinfo{journal}{Nature} \textbf{\bibinfo{volume}{452}},
  \bibinfo{pages}{970} (\bibinfo{year}{2008}).

\bibitem[{\citenamefont{Zhang et~al.}(2009)\citenamefont{Zhang, Liu, Qi, Dai,
  Fang, and Zhang}}]{zhang2009}
\bibinfo{author}{\bibfnamefont{H.}~\bibnamefont{Zhang}},
  \bibinfo{author}{\bibfnamefont{C.-X.} \bibnamefont{Liu}},
  \bibinfo{author}{\bibfnamefont{X.-L.} \bibnamefont{Qi}},
  \bibinfo{author}{\bibfnamefont{X.}~\bibnamefont{Dai}},
  \bibinfo{author}{\bibfnamefont{Z.}~\bibnamefont{Fang}}, \bibnamefont{and}
  \bibinfo{author}{\bibfnamefont{S.-C.} \bibnamefont{Zhang}},
  \bibinfo{journal}{Nature Physics} \textbf{\bibinfo{volume}{5}},
  \bibinfo{pages}{438 } (\bibinfo{year}{2009}).

\bibitem[{\citenamefont{Xia et~al.}(2009)\citenamefont{Xia, Qian, Hsieh, Wray,
  Pal, Lin, Bansil, Grauer, Hor, Cava et~al.}}]{xia2009}
\bibinfo{author}{\bibfnamefont{Y.}~\bibnamefont{Xia}},
  \bibinfo{author}{\bibfnamefont{D.}~\bibnamefont{Qian}},
  \bibinfo{author}{\bibfnamefont{D.}~\bibnamefont{Hsieh}},
  \bibinfo{author}{\bibfnamefont{L.}~\bibnamefont{Wray}},
  \bibinfo{author}{\bibfnamefont{A.}~\bibnamefont{Pal}},
  \bibinfo{author}{\bibfnamefont{H.}~\bibnamefont{Lin}},
  \bibinfo{author}{\bibfnamefont{A.}~\bibnamefont{Bansil}},
  \bibinfo{author}{\bibfnamefont{D.}~\bibnamefont{Grauer}},
  \bibinfo{author}{\bibfnamefont{Y.~S.} \bibnamefont{Hor}},
  \bibinfo{author}{\bibfnamefont{R.~J.} \bibnamefont{Cava}},
  \bibnamefont{et~al.}, \bibinfo{journal}{Nat. Phys.}
  \textbf{\bibinfo{volume}{5}}, \bibinfo{pages}{398} (\bibinfo{year}{2009}).

\bibitem[{\citenamefont{Chen et~al.}(2009)\citenamefont{Chen, Analytis, Chu,
  Liu, Mo, Qi, Zhang, Lu, Dai, Fang et~al.}}]{chen2009}
\bibinfo{author}{\bibfnamefont{Y.~L.} \bibnamefont{Chen}},
  \bibinfo{author}{\bibfnamefont{J.~G.} \bibnamefont{Analytis}},
  \bibinfo{author}{\bibfnamefont{J.~H.} \bibnamefont{Chu}},
  \bibinfo{author}{\bibfnamefont{Z.~K.} \bibnamefont{Liu}},
  \bibinfo{author}{\bibfnamefont{S.~K.} \bibnamefont{Mo}},
  \bibinfo{author}{\bibfnamefont{X.~L.} \bibnamefont{Qi}},
  \bibinfo{author}{\bibfnamefont{H.~J.} \bibnamefont{Zhang}},
  \bibinfo{author}{\bibfnamefont{D.~H.} \bibnamefont{Lu}},
  \bibinfo{author}{\bibfnamefont{X.}~\bibnamefont{Dai}},
  \bibinfo{author}{\bibfnamefont{Z.}~\bibnamefont{Fang}}, \bibnamefont{et~al.},
  \bibinfo{journal}{Science} \textbf{\bibinfo{volume}{325}},
  \bibinfo{pages}{178} (\bibinfo{year}{2009}).

\bibitem[{\citenamefont{Murakami et~al.}(2003)\citenamefont{Murakami, Nagaosa,
  and \textrm{S. C. Zhang}}}]{murakami2003}
\bibinfo{author}{\bibfnamefont{S.}~\bibnamefont{Murakami}},
  \bibinfo{author}{\bibfnamefont{N.}~\bibnamefont{Nagaosa}}, \bibnamefont{and}
  \bibinfo{author}{\bibnamefont{\textrm{S. C. Zhang}}},
  \bibinfo{journal}{Science} \textbf{\bibinfo{volume}{301}},
  \bibinfo{pages}{1348} (\bibinfo{year}{2003}).

\bibitem[{\citenamefont{Qi et~al.}(2008)\citenamefont{Qi, Hughes, and
  Zhang}}]{qi2008}
\bibinfo{author}{\bibfnamefont{X.-L.} \bibnamefont{Qi}},
  \bibinfo{author}{\bibfnamefont{T.}~\bibnamefont{Hughes}}, \bibnamefont{and}
  \bibinfo{author}{\bibfnamefont{S.-C.} \bibnamefont{Zhang}},
  \bibinfo{journal}{Phys. Rev. B} \textbf{\bibinfo{volume}{78}},
  \bibinfo{pages}{195424} (\bibinfo{year}{2008}).

\bibitem[{\citenamefont{Fu et~al.}(2007)\citenamefont{Fu, Kane, and
  Mele}}]{fu2007b}
\bibinfo{author}{\bibfnamefont{L.}~\bibnamefont{Fu}},
  \bibinfo{author}{\bibfnamefont{C.~L.} \bibnamefont{Kane}}, \bibnamefont{and}
  \bibinfo{author}{\bibfnamefont{E.~J.} \bibnamefont{Mele}},
  \bibinfo{journal}{Phys. Rev. Lett.} \textbf{\bibinfo{volume}{98}},
  \bibinfo{eid}{106803} (\bibinfo{year}{2007}).

\bibitem[{\citenamefont{Moore and Balents}(2007)}]{moore2007}
\bibinfo{author}{\bibfnamefont{J.~E.} \bibnamefont{Moore}} \bibnamefont{and}
  \bibinfo{author}{\bibfnamefont{L.}~\bibnamefont{Balents}},
  \bibinfo{journal}{Phys. Rev. B} \textbf{\bibinfo{volume}{75}},
  \bibinfo{eid}{121306} (\bibinfo{year}{2007}).

\bibitem[{wil()}]{wilczek2009}
\bibinfo{note}{F. Wilczek, Nature {\bf 458}, 129 (2009)}.

\bibitem[{\citenamefont{Qi et~al.}(2009)\citenamefont{Qi, Li, Zang, and
  Zhang}}]{qi2009}
\bibinfo{author}{\bibfnamefont{X.-L.} \bibnamefont{Qi}},
  \bibinfo{author}{\bibfnamefont{R.}~\bibnamefont{Li}},
  \bibinfo{author}{\bibfnamefont{J.}~\bibnamefont{Zang}}, \bibnamefont{and}
  \bibinfo{author}{\bibfnamefont{S.-C.} \bibnamefont{Zhang}},
  \bibinfo{journal}{Science} \textbf{\bibinfo{volume}{323}},
  \bibinfo{pages}{1184} (\bibinfo{year}{2009}).

\bibitem[{\citenamefont{Zhang and Hu}(2001)}]{zhang2001}
\bibinfo{author}{\bibfnamefont{S.~C.} \bibnamefont{Zhang}} \bibnamefont{and}
  \bibinfo{author}{\bibfnamefont{J.~P.} \bibnamefont{Hu}},
  \bibinfo{journal}{Science} \textbf{\bibinfo{volume}{294}},
  \bibinfo{pages}{823} (\bibinfo{year}{2001}).

\bibitem[{\citenamefont{Bernevig et~al.}(2002)\citenamefont{Bernevig, Chern,
  Hu, Toumbas, and Zhang}}]{bernevig2002}
\bibinfo{author}{\bibfnamefont{B.~A.} \bibnamefont{Bernevig}},
  \bibinfo{author}{\bibfnamefont{C.~H.} \bibnamefont{Chern}},
  \bibinfo{author}{\bibfnamefont{J.~P.} \bibnamefont{Hu}},
  \bibinfo{author}{\bibfnamefont{N.}~\bibnamefont{Toumbas}}, \bibnamefont{and}
  \bibinfo{author}{\bibfnamefont{S.~C.} \bibnamefont{Zhang}},
  \bibinfo{journal}{Annals of Physics} \textbf{\bibinfo{volume}{300}},
  \bibinfo{pages}{185} (\bibinfo{year}{2002}).

\bibitem[{\citenamefont{Fu and Kane}(2006)}]{fu2006}
\bibinfo{author}{\bibfnamefont{L.}~\bibnamefont{Fu}} \bibnamefont{and}
  \bibinfo{author}{\bibfnamefont{C.~L.} \bibnamefont{Kane}},
  \bibinfo{journal}{Phys. Rev. B} \textbf{\bibinfo{volume}{74}},
  \bibinfo{eid}{195312} (\bibinfo{year}{2006}).

\bibitem[{\citenamefont{Essin et~al.}(2009)\citenamefont{Essin, Moore, and
  Vanderbilt}}]{essin2009}
\bibinfo{author}{\bibfnamefont{A.~M.} \bibnamefont{Essin}},
  \bibinfo{author}{\bibfnamefont{J.~E.} \bibnamefont{Moore}}, \bibnamefont{and}
  \bibinfo{author}{\bibfnamefont{D.}~\bibnamefont{Vanderbilt}},
  \bibinfo{journal}{Phys. Rev. Lett.} \textbf{\bibinfo{volume}{102}},
  \bibinfo{pages}{146805} (\bibinfo{year}{2009}).

\bibitem[{\citenamefont{Dubrovin et~al.}(1985)\citenamefont{Dubrovin, Fomenko,
  and Novikov}}]{dubrovin1985}
\bibinfo{author}{\bibfnamefont{B.~A.} \bibnamefont{Dubrovin}},
  \bibinfo{author}{\bibfnamefont{A.~T.} \bibnamefont{Fomenko}},
  \bibnamefont{and} \bibinfo{author}{\bibfnamefont{S.~P.}
  \bibnamefont{Novikov}}, \emph{\bibinfo{title}{Modern Geometry---Methods and
  Applications, Part 2: The Geometry and Topology of Manifolds}}
  (\bibinfo{publisher}{Springer}, \bibinfo{year}{1985}).

\end{thebibliography}

\end{document}